\documentclass[10pt]{article}
\usepackage[german, english]{babel}
\usepackage[a4paper, left=2.5cm, right=2.5cm, top=3.5cm, bottom=3cm]{geometry}
\usepackage{amsmath}
\usepackage{bbm}
\usepackage{simplewick}
\usepackage{graphicx}
\usepackage{hyperref}
\usepackage{dsfont}
\usepackage{mathrsfs}
\usepackage{color}
\hypersetup{
 colorlinks=true,
 linkcolor=blue,
 citecolor=magenta,
}

\newcommand{\N}{\mathds N}

\newcommand{\A}{\widetilde A}
\newcommand{\R}{\widetilde R}
\newcommand{\rR}{\mathrm{R}}
\newcommand{\cG}{\mathcal{G}}

\def\im{\mathrm{i}}
\def\ep{\mathrm{e}}
\def\pa{\partial}
\def\diff{\mathrm{d}}
\def\tr{\mathrm{tr}}
\def\sfrac#1#2{{\textstyle\frac{#1}{#2}}}
\def\>{\rangle}
\def\<{\langle}
\def\+{\dagger}
\def\we{{\wedge}}
\def\={\ =\ }
\def\unity{\mathbbm{1}}
\def\und{\quad\textrm{and}\quad}
\def\with{\quad\textrm{with}\quad}

\begin{document}

\title{\bf\huge Universal form of the Nicolai map}
\date{~}

\author{\phantom{.}\\[12pt]
{\scshape\Large Olaf Lechtenfeld \ and \ Maximilian Rupprecht}
\\[24pt]
Institut f\"ur Theoretische Physik\\ and\\ 
Riemann Center for Geometry and Physics\\[8pt]
Leibniz Universit\"at Hannover \\ 
Appelstra{\ss}e 2, 30167 Hannover, Germany
\\[24pt]
} 

\clearpage
\maketitle
\thispagestyle{empty}

\begin{abstract}
\noindent\large
The nonlocal bosonic theory obtained from integrating out all anticommuting and
auxiliary variables in a globally supersymmetric theory is characterized by the Nicolai map.
The latter is generated by a coupling flow functional differential operator,
which can be canonically constructed when the supersymmetry is realized off-shell.
Given any scalar superfield theory, we present a universal formula for both the Nicolai map 
and its inverse in terms of an ordered exponential of the integrated coupling flow operator. 
We demonstrate that our formula also holds for supersymmetric gauge theories.
\end{abstract}

\newpage
\setcounter{page}{1} 

\noindent
{\bf Definition of the Nicolai map.\ }
Supersymmetric theories are normally formulated with bosonic and fermionic (and sometimes ghost)
degrees of freedom or, better, using commuting fields~$\phi$ and anticommuting fields~$\psi$. 
Since the latter usually occur only quadratically in the action, they are easily integrated out to produce
a functional determinant. This leaves one with a purely bosonic but nonlocal theory,  given by an action
\begin{equation}
S_g[\phi] \= S^b_g[\phi] + \hbar\,S^f_g[\phi]\ ,
\end{equation}
where $g$ is any coupling constant, and $S^b_g$ and $S^f_g$ denote the local and nonlocal parts
of the new action, respectively. The latter is proportional to the logarithm of the functional determinant 
and is down by a factor of $\hbar$ compared to $S^b_g$, as it generates the fermionic loop contributions 
to expectation values~\footnote{
We work in Minkowski space, but it is also possible to repeat the whole analysis in Euclidean space.}
\begin{equation} \label{pathintegral}
\bigl\< X[\phi] \bigr\>_g \= \int\!{\cal D}\phi\ \exp\bigl\{ \sfrac{\im}{\hbar} S_g[\phi] \bigr\}\ X[\phi]
\end{equation}
of bosonic observables~$X[\phi]$. We assume supersymmetry to be unbroken and non-anomalous, 
so that the vacuum energy vanishes, meaning that $\bigl\<\unity\bigr\>_g=1$ and our expectation values
are already properly normalized. 
We leave the spacetime dimension~$d$ arbitrary and simply write $\diff x$ for its volume element.

In 1980 Hermann Nicolai raised and answered the question of how the nonlocal theory~$S_g$ 
remembers its supersymmetric heritage~\cite{Nic1,Nic2,Nic3} (see also~\cite{JK} for a pedagogical introduction).
Among all such nonlocal bosonic theories, 
the ones originating with a supersymmetric past are characterized by the existence of a
(nonlinear and nonlocal) field transformation (the Nicolai map)
\begin{equation}
T_g :\ \phi(x) \ \mapsto\ \phi'(x;g,\phi)
\end{equation}
invertible at least as a formal power series in~$g$, which admits the key identity~\footnote{
This is not the original definition but an equivalent one as we will show shortly.}
\begin{equation} \label{globalflow}
\bigl\< X[\phi] \bigr\>_g \= \bigl\< X[T_g^{-1}\phi] \bigr\>_0
\qquad\forall\,X \ ,
\end{equation}
relating the interacting theory (at coupling~$g$) to the free one (at coupling~$g{=}0$).
Writing out the path integrals, this requirement is equivalent to
\begin{equation}
{\cal D}\phi\ \exp\bigl\{ \sfrac{\im}{\hbar} S_g[\phi] \bigr\} \=
{\cal D}(T_g\phi)\ \exp\bigl\{ \sfrac{\im}{\hbar} S_0[T_g\phi] \bigr\} \=
{\cal D}\phi\ \exp\bigl\{ \sfrac{\im}{\hbar} S_0[T_g\phi] + \tr\ln\sfrac{\delta T_g\phi}{\delta\phi} \bigr\}\ .
\end{equation}
Collecting powers of $\hbar$, one obtains two conditions,
\begin{equation}
S^b_0[T_g\phi] \= S^b_g[\phi] \quad\und\quad
S^f_0[T_g\phi] -\im\,\tr\ln\sfrac{\delta T_g\phi}{\delta\phi} \= S^f_g[\phi]\ ,
\end{equation}
which originally defined the Nicolai map: 
the local bosonic action is mapped to the free one, and the nonlocal part of the action equals
the Jacobi determinant of the transformation.
We shall henceforth set $\hbar{=}1$ and only use the relation~(\ref{globalflow}) 
to construct the Nicolai map below.

\noindent
{\bf Coupling flow operator.\ }
Except for the rare instances where stochastic variables exist
\cite{Fub1,Fub2,Fub3,FLMR,AFF,Boc1} the Nicolai map 
can only be constructed perturbatively. Therefore, it is reasonable to investigate its infinitesimal version.
This method was developed in~\cite{FL,DL1,L1,DL2,L2}.
Differentiating~(\ref{globalflow}) with respect to the coupling~$g$ yields
\begin{equation} \label{localflow}
\pa_g \bigl\< X[\phi] \bigr\>_g \= \bigl\< \bigl( \pa_g + R_g[\phi] \bigr) X[\phi] \bigr\>_g
\end{equation}
with a functional differential operator
\begin{equation}
R_g[\phi] \= \int\!\diff x\ \bigl(\pa_g T_g^{-1} \circ T_g \bigr) \phi(x)\,\frac{\delta}{\delta\phi(x)}
\end{equation}
that we will refer to as the ``coupling flow operator''. Its knowledge not only guarantees
the existence of the (inverse) Nicolai map but also provides its perturbative construction,
\begin{equation} \label{pertinvT}
\bigl(T_g^{-1} \phi\bigr)(x) \= \exp \Bigl\{ g\,\bigl(\pa_{g'}+R_{g'}[\phi]\bigr)\Bigr\}\ \phi(x) \Bigm|_{g'=0} 
\= \sum_{n=0}^\infty \frac{g^n}{n!}\,\bigl( \pa_{g'} + R_{g'}[\phi] \bigr)^n\ \phi(x) \Bigm|_{g'=0} \ .
\end{equation}
The derivation property of $R_g$ is essential to obtain the distributivity of the (inverse) Nicolai map,
\begin{equation}
T_g^{-1} X[\phi] \= X\bigl[ T_g^{-1}\phi \bigr]\ .
\end{equation}
Alternatively, the map~$T_g$ itself may be found iteratively from the relation~\cite{L2}
\begin{equation} \label{kernel}
\bigl( \pa_g + R_g[\phi]\bigr)\,T_g \phi \= 0
\end{equation}
which immediately follows from~(\ref{globalflow}) for $X=\phi$.

\noindent
{\bf Scalar theories.\ }
How to find the coupling flow operator or at least to show its existence, before knowing~$T_g$?
If the original local theory in terms of $\phi$ and~$\psi$ admits an off-shell supersymmetric formulation,\footnote{
Auxiliary fields may be kept as part of $\phi$ but it is convenient to integrate them out as well.}
then its action~$S_{\textrm{\tiny SUSY}}$ is the highest component of a superfield, 
hence can be expressed as a supervariation~$\delta_\alpha$ of the penultimate component. 
For scalar supermultiplet theories, the same is true for derivatives with respect to the coupling~\cite{FL},
\begin{equation} \label{defDelta}
\pa_g S_{\textrm{\tiny SUSY}}[\phi,\psi] \= \delta_\alpha \Delta_\alpha[\phi,\psi] \ ,
\end{equation}
where $\alpha$ is a Majorana spinor index and $\Delta_\alpha$ is a certain anticommuting functional.\footnote{
For gauge theories, which contain vector supermultiplets, 
the situation is more complicated and will be discussed below.}
Employing~(\ref{defDelta}) and the supersymmetric Ward identity in
\begin{equation}
\pa_g \int\!{\cal D}\phi \int\!{\cal D}\psi\ \exp\bigl\{\im S_{\textrm{\tiny SUSY}}[\phi,\psi] \bigr\}\ X[\phi]
\= \im \int\!{\cal D}\phi \int\!{\cal D}\psi\ \exp\bigl\{\im S_{\textrm{\tiny SUSY}}[\phi,\psi] \bigr\}\
\bigl( \pa_g + \Delta_\alpha[\phi,\psi]\ \delta_\alpha\bigr) X[\phi]\ ,
\end{equation}
we integrate out the anticommuting variables to read off the coupling flow operator
\begin{equation}
R_g[\phi] \= \im\,\bcontraction{}{\Delta}{_\alpha[\phi]\ }{\delta} \Delta_\alpha[\phi]\ \delta_\alpha \=
\im\int\!\diff x\ \bcontraction{}{\Delta}{_\alpha[\phi]\ }{\delta} \Delta_\alpha[\phi]\ \delta_\alpha \phi(x)\ 
\frac{\delta}{\delta\phi(x)}\ ,
\end{equation}
where the contraction indicates the presence of a fermionic propagator obtained from a fermionic bilinear.

The main challenge then is to exponentiate this operator in the construction~(\ref{pertinvT}).
The key new insight here is that the $g'$~derivatives on the right-hand side of~(\ref{pertinvT})
may actually be performed in closed form, by solving a standard differential equation,
\begin{equation}
\pa_g Z_g[\phi] \= Z_g[\phi]\,\bigl( \pa_g + R_g[\phi]\bigr) \qquad\Leftrightarrow\qquad
Z_g[\phi] \= Z_0[\phi]\,\overleftarrow{\cal P} \exp \int_0^g\!\diff h\ R_h[\phi]\ ,
\end{equation}
where $\overleftarrow{\cal P}$ denotes reverse ordering, to be detailled shortly.
With the help of the solution~$Z_g[\phi]$, we obtain
\begin{equation}
\begin{aligned}
T_g^{-1}\phi 
&\= \sum_{n=0}^\infty \frac{g^n}{n!}\,\bigl( Z_{g'}[\phi]^{-1}\,\pa_{g'}\,Z_{g'}[\phi] \bigr)^n\ \phi \Bigm|_{g'=0}
\= \sum_{n=0}^\infty \frac{g^n}{n!}\,Z_{g'}[\phi]^{-1}\,\pa_{g'}^n\,Z_{g'}[\phi]\ \phi \Bigm|_{g'=0} \\[4pt]
&\= Z_{g'}[\phi]^{-1}\;\ep^{g\,\pa_{g'}} Z_{g'}[\phi]\ \phi \Bigm|_{g'=0}
\= Z_{g'}[\phi]^{-1}\,Z_{g'+g}[\phi]\ \phi \Bigm|_{g'=0} \= Z_0[\phi]^{-1}\,Z_g[\phi]\ \phi \ ,
\end{aligned}
\end{equation}
and therefore
\begin{equation} \label{closedinvT}
T_g^{-1}\phi  \= \overleftarrow{\cal P} \exp \Bigl\{ \int_0^g\!\diff h\ R_h[\phi]\Bigr\}\ \phi
\= \sum_{s=0}^\infty \int_0^g\!\diff h_1 \int_0^{h_1}\!\!\!\!\diff h_2 \ldots \int_0^{h_{s-1}}\!\!\!\!\!\!\!\!\!\diff h_s\
R_{h_s}[\phi] \ldots R_{h_2}[\phi]\,R_{h_1}[\phi]\ \phi\ .
\end{equation}
Apparently, the $g'$~derivatives have been traded for integrations, but this representation is more suggestive
than~(\ref{pertinvT}). Moreover, it allows for an immediate formal inversion to write the Nicolai map itself as
\begin{equation} \label{closedT}
T_g\,\phi \= \overrightarrow{\cal P} \exp \Bigl\{-\!\int_0^g\!\diff h\ R_h[\phi]\Bigr\}\ \phi
\= \sum_{s=0}^\infty (-1)^s\!\int_0^g\!\diff h_s \ldots \int_0^{h_3}\!\!\!\!\diff h_2 \int_0^{h_2}\!\!\!\!\diff h_1\
R_{h_s}[\phi] \ldots R_{h_2}[\phi]\,R_{h_1}[\phi]\ \phi\ ,
\end{equation}
with $\overrightarrow{\cal P}$ indicating standard ordering,
This universal form is the main result of our work.

It is instructive to express the power series expansions of $T_g^{-1}\phi$ and of $T_g\phi$ with 
the one for the flow operator,
\begin{equation}
R_g[\phi] \= \sum_{k=1}^\infty g^{k-1} \rR_k[\phi] \= \rR_1[\phi] + g\,\rR_2[\phi] + g^2 \rR_3[\phi] + \ldots
\end{equation}
(the shift in the $g$~power is a practical convention here).
With this, the integrals in (\ref{closedinvT}) and~(\ref{closedT}) can be evaluated to yield
\begin{equation}
T_g^{-1}\phi \= \sum_{\bf n} g^n\,d_{\bf n}\,\rR_{n_s}[\phi]\ldots \rR_{n_2}[\phi]\,\rR_{n_1}[\phi]\ \phi
\end{equation}
and
\begin{equation}
T_g\,\phi \= \sum_{\bf n} g^n\,c_{\bf n}\,\rR_{n_s}[\phi]\ldots \rR_{n_2}[\phi]\,\rR_{n_1}[\phi]\ \phi\ ,
\end{equation}
respectively, where the boldface letter denotes the multiindex
\begin{equation}
{\bf n} = (n_1,n_2,\ldots,n_s) \quad\with n_i\in\N \und \sum_i n_i = n\ ,
\end{equation}
where $1\le s \le n$ and the $n{=}0$ term is the identity.
The numerical coefficients are computed as
\begin{equation}
d_{\bf n} \ = \int_0^1\!\!\diff x_1\;x_1^{n_1-1} \int_0^{x_1}\!\!\!\!\diff x_2\;x_2^{n_2-1} \ldots 
\int_0^{x_{s-1}}\!\!\!\!\!\!\!\!\!\diff x_s\;x_s^{n_s-1} \=
\bigl[ n_s\cdot(n_s+n_{s-1})\cdots(n_s+n_{s-1}+\ldots+n_1)\bigr]^{-1}\ ,
\end{equation}
\begin{equation}
c_{\bf n} \ = (-1)^s\!\int_0^1\!\!\diff x_s\;x_s^{n_s-1} \ldots 
\int_0^{x_3}\!\!\!\!\diff x_2\;x_2^{n_2-1} \int_0^{x_2}\!\!\!\!\diff x_1\;x_1^{n_1-1} \=
(-1)^s\bigl[ n_1\cdot(n_1+n_2)\cdots(n_1+n_2+\ldots+n_s)\bigr]^{-1} \ ,
\end{equation}
the latter being in agreement with the earlier result in~\cite{L2} derived from~(\ref{kernel}).
Writing out the first few terms, the perturbative Nicolai map reads
\begin{equation}
\begin{aligned}
T_g\phi &\= \phi \ -\ g\,\rR_1 \phi \ -\ \sfrac12g^2\bigl(\rR_2-\rR_1^2\bigr)\phi\ -\ 
\sfrac16g^3\bigl(2\rR_3-\rR_1\rR_2-2\rR_2\rR_1+\rR_1^3\bigr)\phi \\
&\quad -\sfrac{1}{24}g^4\bigl(6\rR_4-2\rR_1\rR_3-3\rR_2\rR_2+\rR_1^2\rR_2-6\rR_3\rR_1
+2\rR_1\rR_2\rR_1+3\rR_2\rR_1^2-\rR_1^4\bigr)\phi \ +\ {\cal O}(g^5)\, .
\end{aligned}
\end{equation}

\noindent
{\bf Gauge theories.\ }
Supersymmetric Yang--Mills theory is a cornerstone of modern mathematical physics and therefore
of prime interest. The Nicolai map promises an alternative approach to its quantization and has
regained some attention recently~\cite{ANPP,NP,ALMNPP,AMPP}.
Let us hence consider unbroken ${\cal N}{=}\,1$ supersymmetric gauge theories in the Wess--Zumino gauge,
with the field content $(A,\lambda,D)$ in the adjoint representation of the gauge group 
and the Yang--Mills field strength
\begin{equation}
F \= \diff A + g\,A\we A \= \sfrac12 F_{\mu\nu}\,\diff x^\mu\we\diff x^\nu \ .
\end{equation}
Choosing a linear gauge fixing~\footnote{
for convenience; nonlinear gauges are easily accomodated 
with $\cG(A)=\tilde\cG(gA)/g$ for an arbitrary function~$\tilde\cG$.}
\begin{equation}
0 \= \cG(A) \= \pa^\mu\! A_\mu \quad\textrm{or}\quad n^\mu\! A_\mu
\end{equation}
adds, via the Faddeev--Popov trick and the 't Hooft averaging, a gauge-fixing term 
depending on ghost fields $C$ and~$\bar{C}$ and a gauge parameter~$\xi$ to the action. 
This explicitly breaks supersymmetry and reduces the gauge symmetry to BRST invariance.
The construction of the Nicolai map now presents an additional challenge, 
because the $g$~derivative of 
\begin{equation}
S_{\textrm{\tiny SUSY}}[A,\lambda,D,C,\bar{C}] \= 
\int\!\diff x\ \tr\Bigl\{ -\sfrac14 F_{\mu\nu}F^{\mu\nu} - \sfrac{1}{2\xi}\cG(A)^2 
+ \textrm{fermions} + \textrm{ghosts} + \textrm{auxiliaries} \Bigr\}
\end{equation}
is no longer a supervariation, even not up to a Slavnov variation (which generates the BRST transformations). 
The way out is a rescaling of all fields with a suitable power of~$g$, with the total Jacobian being unity. 
In particular, for the commuting fields we define
\begin{equation} \label{rescale}
\A = g\,A \qquad\Rightarrow\qquad \widetilde{F} = g\,F = \diff\A + \A\we\A
= \sfrac12 \widetilde{F}_{\mu\nu}\,\diff x^\mu\we\diff x^\nu 
\quad\und\quad \widetilde{D} = g\,D
\end{equation}
and arrive at
\begin{equation}
S_{\textrm{\tiny SUSY}}[\A,\widetilde{\lambda},\widetilde{D},\widetilde{C},\widetilde{\bar{C}}] \= 
\frac{1}{g^2}\int\!\diff x\ \tr\Bigl\{ -\sfrac14 \widetilde{F}_{\mu\nu}\widetilde{F}^{\mu\nu} 
- \sfrac{1}{2\xi}\cG(A)^2 + \textrm{fermions} + \textrm{ghosts} + \textrm{auxiliaries} \Bigr\}\ ,
\end{equation}
where the only $g$~dependence resides in front of the integral and in a factor of $g$ multiplying the ghost term.
It is then easy to show that~\cite{DL1,L1}
\begin{equation}
\pa_g S_{\textrm{\tiny SUSY}} \= -\frac{1}{g^3}\,\Bigl\{ 
\delta_\alpha \Delta_\alpha[\A,\widetilde{\lambda},\widetilde{D}] 
- \sqrt{g}\,s\,\Delta_{\textrm{gh}}[\widetilde{\bar{C}},\A] \Bigr\}
\quad\with\quad \Delta_{\textrm{gh}} = \smallint\tr\bigl\{\widetilde{\bar{C}}\,\cG(\A)\bigr\}
\end{equation}
where $\Delta_\alpha$ is a particular gauge-invariant fermionic functional 
and `$s$' denotes the (anticommuting) Slavnov variation.
With this information, we can employ the Ward identities for BRST and broken supersymmetry to compute
the effect of a $g$~derivative on the expectation value $\bigl\<X[\A]\bigr\>_g$ after integrating out
gaugini, ghosts and auxiliaries,
\begin{equation}
\pa_g \bigl\< X[\A] \bigr\>_g \= \bigl\< \bigl( \pa_g + \sfrac1g\R[\A] \bigr) X[\A] \bigr\>_g
\end{equation}
where the coupling flow operator is given by~\cite{DL1,L1}
\begin{equation} \label{gaugeR}
\R[\A] \= -\im\,\bcontraction{}{\Delta}{_\alpha[\A]\ }{\delta} \Delta_\alpha[\A]\ \delta_\alpha 
+\sfrac{\im}{\sqrt{g}}\,\bcontraction{}{\Delta}{_{\textrm{gh}}[\A]\ }{s} \Delta_{\textrm{gh}}[\A]\ s
-\sfrac{1}{\sqrt{g}}\,\bcontraction{}{\Delta}{_\alpha[\A]\ \bigl(}{\delta}  \Delta_\alpha[\A]\ \bigl(\delta_\alpha 
\bcontraction{}{\Delta}{_{\textrm{gh}}[\A]\bigr)\ }{s} \Delta_{\textrm{gh}}[\A]\bigr)\ s
\quad\with\quad s\,\A_\mu = \sqrt{g}\,\widetilde{D}_\mu\widetilde{C}
\end{equation}
and contractions indicating either gaugino or ghost propagators.
It is just a complicated linear functional differential operator.
Remarkably, in the variable~$\A$ it is independent of the gauge coupling~$g$.

Observing that 
$\bcontraction{}{\Delta}{_{\textrm{gh}}[\A]\ }{s} \Delta_{\textrm{gh}}[\A]\ s\,\cG(\A) = -\im\sqrt{g}\,\cG(\A)$,
it follows that
\begin{equation}
\R[\A]\,\cG(\A) \= \cG(\A) \qquad\Rightarrow\qquad
\bigl( \pa_g + \sfrac1g \R[\A] \bigr)\,\sfrac1g\,\cG(\A) \= 0\ ,
\end{equation}
which implies that the chosen gauge class $\cG(A)=\sfrac1g\,\cG(\A)$ is invariant 
under the coupling constant flow generated by~$\R$ and hence a fixed point of the Nicolai map. 
This property is an additional requirement in the usual definition of the Nicolai map~\cite{Nic3}, 
but it is automatic here from the construction of~$\R$.

In order to integrate the coupling flow and obtain the analog of~(\ref{pertinvT})
it is necessary to revert the rescaling~(\ref{rescale}),
\begin{equation}
T_g^{-1} A \= \sfrac1g\,\exp \Bigl\{ g\,\bigl(\pa_{g'}+\sfrac{1}{g'}\R[\A]\bigr)\Bigr\}
\ \A \Bigm|_{\A=g'\!A}\,\Bigm|_{g'=0} \ .
\end{equation}
It is not manifest but true that the final step $g'\to0$ is nonsingular.
The $g'$ derivatives in the exponent can be executed,
\begin{equation}
\begin{aligned}
T_g^{-1} A &\= \frac1g\,\sum_{n=0}^\infty \frac{g^n}{n!}\,
\bigl(\pa_{g'}+\sfrac{1}{g'}\R[\A]\bigr)^n \ \A \Bigm|_{\A=g'\!A}\,\Bigm|_{g'=0} \\[4pt]
&\= \frac1g\,\sum_{n=0}^\infty \frac{g^n}{n!}\,
\Bigl( (g')^{-\R[\A]}\,\pa_{g'}\,(g')^{\R[\A]} \Bigr)^n \ \A \Bigm|_{\A=g'\!A}\,\Bigm|_{g'=0} \\[4pt]
&\= \frac1g\,\sum_{n=0}^\infty \frac{g^n}{n!}\,
(g')^{-\R[\A]}\,\pa_{g'}^n\,(g')^{\R[\A]} \ \A \Bigm|_{\A=g'\!A}\,\Bigm|_{g'=0} \\[4pt]
&\= \frac1g\, (g')^{-\R[\A]}\,\exp\bigl\{ g\,\pa_{g'}\bigr\}\,(g')^{\R[\A]} \ \A \Bigm|_{\A=g'\!A}\,\Bigm|_{g'=0} \\[4pt]
&\= \frac1g\, (g')^{-\R[\A]}\,(g'+g)^{\R[\A]} \ \A \Bigm|_{\A=g'\!A}\,\Bigm|_{g'=0} 
\ \=\ \frac{g'}{g}\,\Bigl( 1+\frac{g}{g'} \Bigr)^{\R[g'\!A]} \ A \Bigm|_{g'=0} \\[4pt]
&\= \sum_{n=1}^\infty \frac{1}{n!}\,\Bigl(\frac{g}{g'}\Bigr)^{n-1}\,\R[g'\!A]\,\bigl(\R[g'\!A]-1\bigr)
\cdots\bigl(\R[g'\!A]-n{+}1\bigr)\ A \Bigm|_{g'=0} \ ,
\end{aligned}
\end{equation}
but regularity at $g'=0$ is still not obvious. In order to clarify this property, 
it is convenient to break up~$\R$ into homogeneous pieces and split off the degree-zero part
(remember $\A=g\,A$),
\begin{equation}
\R[\A] \= \smash{\sum_{k=0}^\infty}\,\rR_k[\A] \ =:\ \rR_0[A]+g\,R_g[A] \quad\with\quad
E\,\rR_k[\A]\ \equiv\ \smallint\!\A\,\sfrac{\delta}{\delta\A}\ \rR_k[\A] \= k\,\rR_k[\A]\ ,
\end{equation}
where we defined the functional Euler operator~$E$.
When scaling back from $\A$ to~$A$, it is useful to recall for any functional~$F$ the obvious equivalence
\begin{equation}
g\,\pa_g\, F[\A] = 0 \qquad\Leftrightarrow\qquad
\bigl( g\,\pa_g - E \bigr)\,F[gA] = 0 \ .
\end{equation}
Applying this in the fourth line below and twice employing $\bigl[g\pa_g,\frac1g\bigr]=-\frac1g$, we obtain
\begin{equation}
\begin{aligned}
T_g^{-1} A &\= \frac1g\,\sum_{n=0}^\infty \frac{g^n}{n!}\,\Bigl[\frac{1}{g'}\,
\bigl( g'\pa_{g'} + \R[\A] \bigr)\Bigr]^n\ \A \Bigm|_{\A=g'\!A}\,\Bigm|_{g'=0}\\[2pt]
&\=  \frac1g\,\sum_{n=1}^\infty \frac{1}{n!}\Bigl(\frac{g}{g'}\Bigr)^n\,\bigl( \R[\A]-n{+}1 \bigr)
\cdots \bigl( \R[\A]-1 \bigr)\,\R[\A]\ \A \Bigm|_{\A=g'\!A}\,\Bigm|_{g'=0}\\[2pt]
&\= \frac1g\,\sum_{n=1}^\infty \frac{1}{n!}\Bigl(\frac{g}{g'}\Bigr)^n\,\bigl( \R[g'\!A]-n{+}1 \bigr)
\cdots \bigl( \R[g'\!A]-1\bigr)\,\R[g'\!A]\ g'\!A \,\Bigm|_{g'=0}\\[2pt]
&\= \frac1g\,\sum_{n=1}^\infty \frac{1}{n!}\Bigl(\frac{g}{g'}\Bigr)^n\,\bigl( g'\pa_{g'}{-}E+\R[g'\!A]-n{+}1 \bigr)
\cdots \bigl( g'\pa_{g'}{-}E+\R[g'\!A]\bigr)\ g'\!A \,\Bigm|_{g'=0}\\[2pt]
&\= \frac1g\,\sum_{n=1}^\infty \frac{g^n}{n!}\,\Bigl[\frac{1}{g'}\,
\bigl( g'\pa_{g'} -E + \R[g'\!A] \bigr)\Bigr]^n\ g'\!A \,\Bigm|_{g'=0}\\[2pt]
&\= \frac1g\,\sum_{n=1}^\infty \frac{g^n}{n!}\, 
\bigl( \pa_{g'} +\sfrac{1}{g'}(\rR_0[A]{-}E) + R_{g'}[A] \bigr)^n\ g'\!A \,\Bigm|_{g'=0}\\[2pt]
&\= \frac1g\,\sum_{n=1}^\infty \frac{g^n}{n!}\, \bigl( \pa_{g'} + R_{g'}[A] \bigr)^n\ g'\!A \,\Bigm|_{g'=0}
\ \=\ \frac1g\,\sum_{n=1}^\infty \frac{g^n}{n!}\,n\,\bigl( \pa_{g'} + R_{g'}[A] \bigr)^{n-1}\ A \,\Bigm|_{g'=0}\ ,
\end{aligned}
\end{equation}
where we noticed and used the necessity
\begin{equation}
\rR_0[A] = E = \smallint\! A\,\sfrac{\delta}{\delta A} \qquad\Leftrightarrow\qquad
T_g^{-1} A \= A + g\,\rR_1[A]\,A +\, {\cal O}(g^2)\ ,
\end{equation}
which is borne out by explicit computation as well~\cite{L2,ALMNPP}. 
Therefore, with the coupling flow operators in the two field scalings being related by
\begin{equation} \label{Rrel}
R_g[A] \= \sfrac1g\,\bigl( \R[gA] -E \bigr)\ ,
\end{equation}
the final inverse Nicolai map reads
\begin{equation} \label{pertinvTA}
T_g^{-1} A \= \sum_{n=0}^\infty \frac{g^n}{n!}\,\bigl( \pa_{g'} + R_{g'}[A] \bigr)^n\ A \,\Bigm|_{g'=0}
\ \=\ \exp\Bigl\{ g\,\bigl(\pa_{g'}+R_{g'}[A]\bigr)\Bigr\}\ A \,\Bigm|_{g'=0}\ .
\end{equation}
This form has been employed directly already in~\cite{Nic2,Nic3} for $d{=}4$ and again in~\cite{ALMNPP}
for the critical dimensions $d{=}3,4,6$ and~$10$. Appendix~A of~\cite{ALMNPP} generalized the earlier
proof of existence to all critical dimensions without the need for off-shell supersymmetry, 
but only in the Landau gauge.

Obviously, (\ref{pertinvTA}) is of the same form as~(\ref{pertinvT}) for scalar theories.
Therefore, the universal forms (\ref{closedinvT}) and (\ref{closedT}) apply for gauge theories as well,
\begin{equation}
T_g A \= \overrightarrow{\cal P} \exp \Bigl\{-\!\int_0^g\!\diff h\ R_h[A]\Bigr\}\ A
\= \sum_{\bf n} g^n\,c_{\bf n}\,\rR_{n_s}[A]\ldots \rR_{n_2}[A]\,\rR_{n_1}[A]\ A\ .
\end{equation}
This, together with (\ref{Rrel}), is our second main result.

\noindent
{\bf Conclusions and outlook.\ }
The property $\bigl\< X[\phi] \bigr\>_g=\bigl\< X[T_g^{-1}\phi] \bigr\>_0$ 
suffices to define the inverse Nicolai map~$T_g^{-1}$. 
We briefly reviewed how off-shell supersymmetry admits the construction of a coupling flow operator~$R_g[\phi]$,
which generates the inverse Nicolai map via exponentiation of~$\pa_g+R_g[\phi]$. 
The $g$~derivatives can be integrated to find a universal formula for the Nicolai map as 
$T_g=\overrightarrow{\cal P}\exp\bigl\{-\int_0^g\diff h\,R_h[\phi]\bigr\}$
given by an ordered exponential. 
This formula applies both to scalar and gauge superfield theories, 
and it recovers the correct power series expansion of~$T_g$. 
For gauge theories the coupling flow automatically respects the gauge choice.

Various applications are in sight, namely 
the gauge dependence and uniqueness of the Nicolai map, 
the absence of off-shell supersymmetry in higher dimensions, 
nonlinear sigma models, 
extended supersymmetry,
or supersymmetry breaking
(which may be triggered by an external field, 
as for the matrix models in~\cite{KurSug}).
Since Lorentz invariance is not necessary for our construction, 
our scope includes non-Lorentzian theories with off-shell supersymmetry, 
such as~\cite{galilean}.
We hope to come back to these issues.

\bigskip

\noindent
{\bf Acknowledgment.\ } 
M.R.~is supported by a PhD grant of the German Academic Scholarship Foundation.


\newpage

\begin{thebibliography}{99}

\bibitem{Nic1}
H.~Nicolai,
{\it On a new characterization of scalar supersymmetric theories},\\
\href{https://dx.doi.org/10.1016/0370-2693(80)90138-0}
{{\it Phys.\ Lett.\ B} {\bf 89} (1980) 341}.
\bibitem{Nic2}
H.~Nicolai,
{\it Supersymmetry and functional integration measures},\\
\href{https://dx.doi.org/10.1016/0550-3213(80)90460-5}
{{\it Nucl.\ Phys.\ B} {\bf176} (1980) 419}.
\bibitem{Nic3}
H.~Nicolai,
{\it Supersymmetric functional integration measures},\\
lectures delivered at the NATO Advanced Study Institute on Supersymmetry,\\
Bonn, Germany, 20--31 Aug 1984, 
\href{https://cds.cern.ch/record/155731?ln=en}
{pp.393--420, eds. K.~Dietz et. al., {\it Plenum Press} (1984)}.
\bibitem{JK}
H.~Ezawa and J.~Klauder,
{\it Fermions without fermions: the Nicolai map revisited}, \\
\href{https://dx.doi.org/10.1143/PTP.74.904}
{{\it Prog.\ Theor.\ Phys.} {\bf 74} (1985) 904}.
\bibitem{Fub1}
V.~de Alfaro, S.~Fubini, G.~Furlan and G.~Veneziano,\\
{\it Stochastic identities in supersymmetric theories},
\href{https://dx.doi.org/10.1016/0370-2693(84)91349-2}
{{\it Phys.\ Lett.\ B} {\bf 142} (1984) 389}.
\bibitem{Fub2}
V.~de Alfaro, S.~Fubini, G.~Furlan and G.~Veneziano,\\
{\it Stochastic identities in quantum theory},
\href{https://dx.doi.org/10.1016/0370-2693(85)90215-1}
{{\it Nucl.\ Phys.\ B} {\bf 255} (1985) 1}.
\bibitem{Fub3}
V.~de Alfaro, S.~Fubini, G.~Furlan and G.~Veneziano,
{\it Nicolai mapping and stochastic identities in supersymmetric field theories},
\href{https://dx.doi.org/10.1016/0370-1573(86)90071-2}
{{\it Phys.\ Rept.} {\bf 137} (1986) 55}.
\bibitem{FLMR}
R.~Floreani, J.P.~Leroy, J.~Michel and G.C.~Rossi,\\
{\it A perturbative study of the Nicolai mapping},
\href{https://dx.doi.org/10.1016/0370-2693(85)90736-1}
{{\it Phys.\ Lett.\ B} {\bf 158} (1985) 47}.
\bibitem{AFF}
V.~de Alfaro, S.~Fubini and G.~Furlan,
{\it Stochastic identities in the light cone gauge},\\
\href{https://dx.doi.org/10.1016/0370-2693(85)90215-1}
{{\it Phys.\ Lett.\ B} {\bf 163} (1985) 176}.
\bibitem{Boc1}
M.~Bochicchio and A.~Pilloni,
{\it Gauge theories in anti-selfdual variables},\\
\href{https://dx.doi.org/10.1007/JHEP09(2013)039}
{{\it JHEP} {\bf 09} (2013) 039}
[\href{https://arxiv.org/abs/1304.4949}{arXiv:1304.4949 [hep-th]}].
\bibitem{FL}
R.~Flume and O.~Lechtenfeld,
{\it On the stochastic structure of globally supersymmetric field theories},\\
\href{https://dx.doi.org/10.1016/0370-2693(84)90459-3}
{{\it Phys.\ Lett.\ B} {\bf 135} (1984) 91}.
\bibitem{DL1}
K.~Dietz and O.~Lechtenfeld,
{\it Nicolai maps and stochastic observables from a coupling constant flow},\\
\href{https://dx.doi.org/10.1016/0550-3213(85)90132-4}
{{\it Nucl.\ Phys.\ B} {\bf 255} (1985) 149}.
\bibitem{L1}
O.~Lechtenfeld,
{\it Construction of the Nicolai mapping in supersymmetric field theories},\\
Ph.D.\ Thesis, Bonn University (1984),
\href{https://lib-extopc.kek.jp/preprints/PDF/2000/0030/0030157.pdf}
{internal report {\it BONN-IR-84-42}, ISSN-0172-8741}.
\bibitem{DL2}
K.~Dietz and O.~Lechtenfeld,
{\it Ghost-free quantisation of non-Abelian gauge theories 
via the Nicolai transformation of their supersymmetric extensions},
\href{https://dx.doi.org/10.1016/0550-3213(85)90642-X}
{{\it Nucl.\ Phys.\ B} {\bf 259} (1985) 397}.
\bibitem{L2}
O.~Lechtenfeld, 
{\it Stochastic variables in ten dimensions?},
\href{https://dx.doi.org/10.1016/0550-3213(86)90531-6}
{{\it Nucl.\ Phys.\ B} {\bf 274} (1986) 633}.
\bibitem{ANPP}
S.~Ananth, H.~Nicolai, C.~Pandey and S.~Pant,
{\it Supersymmetric Yang--Mills theories: not quite the usual perspective},
\href{httsp://dx.doi.org/10.1088/1751-8121/ab7b9d}
{{\it J.\ Phys.\ A} {\bf 53} (2020) 174001}
[\href{https://arxiv.org/abs/2001.02768}{arXiv:2001.02768 [hep-th]}].
\bibitem{NP}
H.~Nicolai and J.~Plefka,
{\it ${\cal N}{=}\,4$ super-Yang--Mills correlators without anticommuting variables},\\
\href{https://dx.doi.org/10.1103/PhysRevD.101.125013}
{{\it Phys.\ Rev.\ D} {\bf 101} (2020) 125013}
[\href{https://arxiv.org/abs/2003.14325}{arXiv:2003.14325 [hep-th]}].
\bibitem{ALMNPP}
S.~Ananth, O.~Lechtenfeld, H.~Malcha, H.~Nicolai, C.~Pandey and S.~Pant,
{\it Perturbative linearization of supersymmetric Yang--Mills theory},
\href{https://dx.doi.org/10.1007/JHEP10(2020)199}
{{\it JHEP} {\bf 10} (2020) 199}
[\href{https://arxiv.org/abs/2005.12324}{arXiv:2005.12324 [hep-th]}].
\bibitem{AMPP}
S.~Ananth, H.~Malcha, C.~Panday and A.~Pant,
{\it Supersymmetric Yang--Mills theory in $D{=}6$ without anticommuting variables},
\href{https://dx.doi.org/10.1103/PhysRevD.103.025010}
{{\it Phys.\ Rev.\ D} {\bf 103} (2021) 025010} 
[\href{https://arxiv.org/abs/2006.02457}{arXiv:2006.02457 [hep-th]}].
\bibitem{KurSug}
T.~Kuroki and F.~Sugino,
{\it Spontaneous supersymmetry breaking in matrix models 
from the viewpoints of localization and Nicolai mapping},
\href{https://dx.doi.org/10.1016/j.nuclphysb.2010.11.015}
{\it Nucl.\ Phys.\ B {\bf 844} (2011) 409}
[\href{https://arxiv.org/abs/1009.6097}{arXiv:1009.6097 [hep-th]}].
\bibitem{galilean}
R.~Auzzi, S.~Baiguera, G.~Nardelli and S.~Penati,
{\it Renormalization properties of a Galilean Wess--Zumino model},
\href{https://dx.doi.org/10.1007/JHEP06(2019)048}
{{\it JHEP} {\bf 06} (2019) 048}
[\href{https://arxiv.org/abs/2005.12324}{arXiv:2005.12324 [hep-th]}].

\end{thebibliography}
\end{document}